\documentclass[twocolumn,superscriptaddress,altaffilletter,prd]{revtex4-1}

\usepackage[margin=1.0in]{geometry}			
\usepackage{lineno}							
\usepackage{titletoc}						
\usepackage{color}							
\usepackage{graphicx}						
\usepackage{hyperref}						
\usepackage{fancyhdr}						
\usepackage{enumitem}						
\usepackage{url}							
\usepackage{siunitx}						
\usepackage[breakable, theorems, skins]{tcolorbox}	
\usepackage{array}							
\usepackage{longtable}                      
\usepackage{multirow}                       
\usepackage[title]{appendix}                
\newcommand{\iso}[2]{$^{#1}$#2}

\begin{document}

%
%

\title{Using Cyclotron Radiation Emission for Ultra-high Resolution X-Ray Spectroscopy}

\author{K. Kazkaz}
\email[Correspondence to: ]{kareem@llnl.gov}
\affiliation{Lawrence Livermore National Laboratory, Livermore, CA 94550, USA}
\author{N. Woollett}
\affiliation{Lawrence Livermore National Laboratory, Livermore, CA 94550, USA}

\date{\today}

\begin{abstract}
Cyclotron Radiation Emission Spectroscopy (CRES) is an approach to measuring the energy of an electron trapped in an externally applied magnetic field. The bare electron can come from different interactions, including photoelectric absorption, Compton scatters, beta decay, and pair production. CRES relies on measuring the frequency of the electron's cyclotron motion, and because the measurement times extend over $10^6$-$10^7$ cycles, the energy resolution is on the order of a single electronvolt. To date, CRES has only been performed on internal beta-emitting radioisotopes, but the technology can be applied to X-ray spectrometery through appropriate selection of a target gas and sufficient intensity of the distinct X-ray source. The applications of this technology range from high-precision measurements of atomic energy levels, to calibrations of basic science experiments, to trace element identification. In this work we explore the use of CRES for X-ray spectroscopy within the rubric of measuring the energy levels of argon, although the principles are broadly applicable to many other situations. The issues we explore include target material, density, electron trapping depth, noise levels, and overall efficiency. We also discuss spectral deconvolution and how the multiple peaks obtained from a single target / source pair can be used to enhance the robustness of the measurement.


\end{abstract}

\maketitle

%
%

\section{Introduction}
\label{s:Intro}

Throughout the history of particle physics, several key physical processes have been exploited in the construction of radiation detectors with the most prevalent technologies relying on scintillation, ionization, calorimetry, and Cherenkov production.
Each fundamental approach comes with both advantages and disadvantages when compared through established metrics such as resolution, intrinsic efficiency, timing, pulse shape discrimination capability, required infrastructure, volume usage, and cost.
No one detector or underlying technology is suited to all situations, and some detector technologies make use of more than one underlying process to enhance performance.
It is therefore important to recognise new developments which could lead to substantial improvements in the performance metrics.

One of the metrics of greatest interest is detector resolution.
For X-ray measurements, the highest-resolution microcalorimeters can have better than 0.1\% resolution~\cite{Porter,Ullom}.
Diffraction systems can have resolution on the order of 1 part in $10^7$~\cite{Masciovecchio}.
These instruments, however, obtain such levels of performance over a narrow energy range, of order 10's of keV for microcalorimeters, while for an X-ray diffraction system the energy range is of order 10\,meV, and outside this range the diffraction system must be re-tuned for a different energy.
For more general-purpose, broad-band semiconductor detectors, representative energy resolution is on the order of 1\%~\cite{Wiacek}. Measurements enabled by ultrahigh resolution X-ray detection range from astrophysics~\cite{Kuhn} to ultrafast X-ray measurements~\cite{Miaja-Avila} to nonlinear optical effects~\cite{Rohringer,Tamasaku}, to name but a few.

A relatively new method for measuring electron energies is via cyclotron radiation emission spectroscopy, or CRES.
Originally proposed in 2009~\cite{Monreal} and proven feasible in 2015~\cite{Asner}, its original purpose was to perform a direct measurement of the mass of the electron anti-neutrino via a high-resolution analysis of the endpoint of the atomic tritium beta spectrum.
The prototype detector obtained an instrumental resolution of 2.8\,eV~\cite{Esfahani}, with an  optimized system perhaps capable of sub-eV resolution~\cite{Robertson}.
Unlike microcalorimeters and X-ray diffraction energy measurements, this resolution is anticipated to be largely independent of energy, allowing for 0.1\% at 1\,keV to 0.001\% for transuranic K-shell transitions.

The effective energy range of the simplest CRES system is on the order of 10\,keV (e.g., an effective range of 15-25 keV, or 70-80 keV, or any other range where the minimum and maximum are separated by 10 keV), although separating and amplifying the signal line can increase this range to over 100\,keV via use of multiple down-mixers (see Section~\ref{s:ExampleCase}).
Because a single system would be used to probe multiple targets and sources, both absolute and relative measurements could be obtained, with the systematic uncertainties of the latter category greatly reduced compared to the nominal resolution and efficiency of the detector.
We focus on noble element targets, as energy sharing with an extended molecule will result in a systematically broadened energy peak.
These measurements would provide a valuable test of the precision of theoretical calculations which are significant within multiple branches of physics.

In this paper we outline a general purpose CRES detector capable of detecting X-rays with high energy resolution.
In Section~\ref{s:Argon} we use the measurement of argon energy levels to motivate the exploration of the technical issues surrounding the use of CRES for X-ray spectroscopy. We discuss a possible apparatus in Section~\ref{s:Apparatus}, and an exploration of the anticipated principles and interactions that inform the detailed set-points of the detector in Section~\ref{s:System Performance}.
Then in Section~\ref{s:ExampleCase} we apply the principles to a system optimized for measuring the atomic energy levels of argon and its photoelectric cross-section with 6-keV X-rays.
Section~\ref{s:ExampleCase} also contains a description of the spectral deconvolution required to reconstruct the original X-ray spectrum.
We end with a discussion of considerations beyond the example argon system.

\section{Measuring argon energy levels}
\label{s:Argon}

We motivate our exploration of the experimental choices and set points of CRES detectors via measurement of the energy levels of the bound electrons in a ground-state argon atom.
Modern physics tools and databases involving the X-ray absorption edges for the elements~\cite{XCOM,KayeLaby,Geant4,MCNP} are largely based on a small number of publications that are half a century old~\cite{Scofield,Bearden}.
During the intervening time, these databases have been cross-compared with both theoretical~\cite{Pratt} and experimental~\cite{Saloman} approaches, though a full review of validations of the databases is beyond the scope of this work.

These cross-checks, however, can have systematic limitations based on disagreements in the underlying theory.
For example, the NIST XCOM databases have discrepancies in the interaction cross-sections of up to 15\% far from the absorption edges, and over 50\% near them.
On the experimental side, the comparisons are limited by the instrumental resolution of the X-ray detector.


In our treatment of a CRES-based X-ray spectrometer, we will focus primarily on the instrumental effects and resolution. We remain mindful, however, of physics-based limitations to final resolution based on, e.g., Doppler shifting or the lifetime of the excited state leading to natural peak broadening. A CRES system may therefore exhibit larger resolution than anticipated solely from the treatment in the current work, but the amount of broadening will vary from case to case, depending on the underlying physics.

%
%

\section{Apparatus}
\label{s:Apparatus}

This section is a discussion of the considerations for the design of a general CRES X-ray spectrometer. The RF design and subsequent signal analysis are the major guiding factors.

\begin{figure}[t]
\includegraphics[width=0.9\columnwidth]{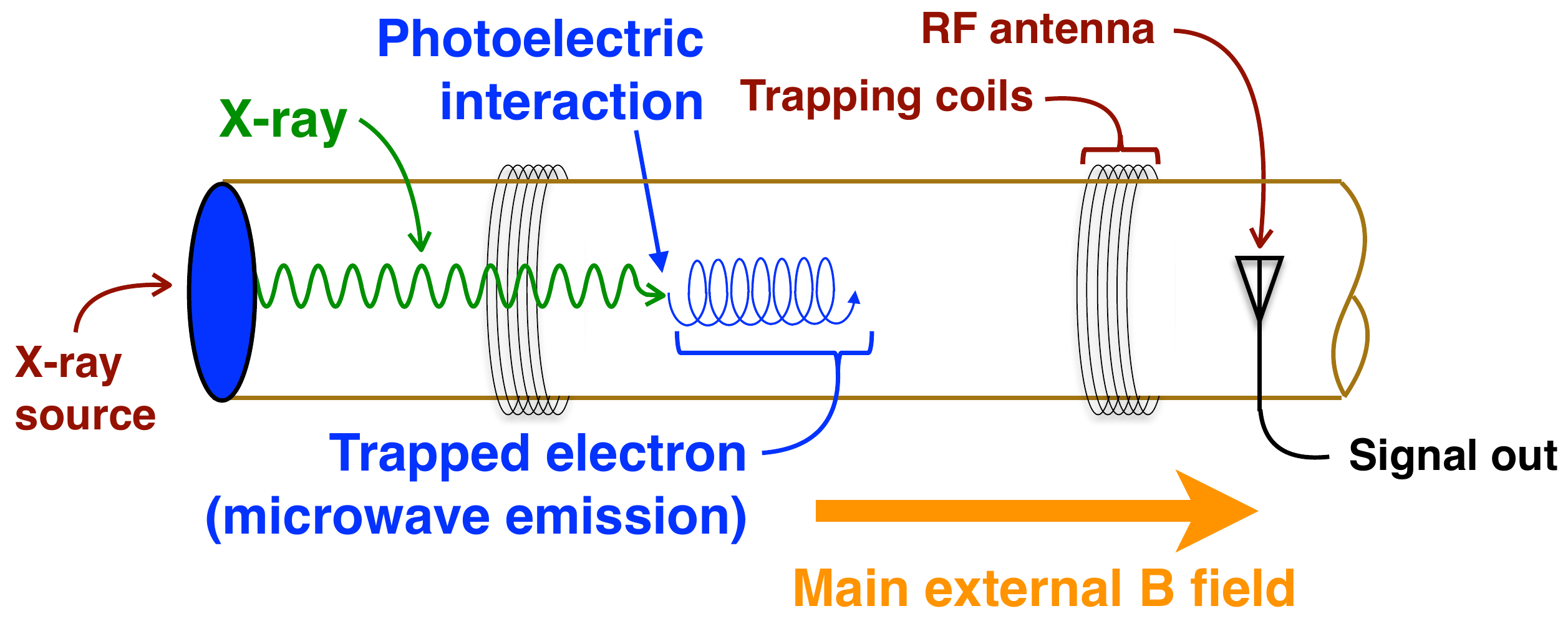}
\caption{Schematic of CRES system optimized for X-ray detection. The depicted X-ray source can be either a naturally occurring X-ray source such as \iso{55}{Fe} or \iso{49}{V}. Alternatively, the source can be a low-density window that permits introduction of externally generated X-rays. The entire apparatus is embedded in a main Tesla-scale magnetic field to induce the cyclotron motion.}
\label{fig:schematic}
\end{figure}

To build an X-ray detector using CRES principles, we rely on photoelectric interactions between the X-ray and a target atom.
The target is a low-pressure gas such that the mean free path of the electron will allow on the order of hundreds of microseconds to  milliseconds of flight time to allow for track reconstruction.
The target gas is held in a Tesla-scale magnetic field that causes ejected electrons to undergo cyclotron motion, emitting microwave radiation.
Magnetic trapping coils then keep the electron confined to an instrumented volume (Fig.\,\ref{fig:schematic}), allowing for long-term collection of the radiated microwaves without the need for large target volume.
The signal from the microwave antenna is amplified and digitized, and a Fourier transform applied to create a power spectrum for any given time segment.

\begin{figure}[t]
\includegraphics[width=\columnwidth]{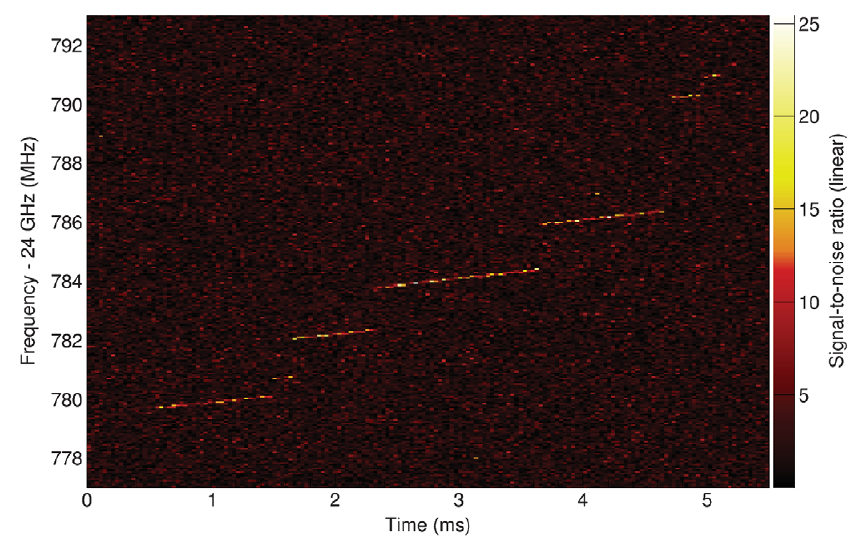}
\caption{Fourier transform of the signal from a trapped electron from Ref.~\cite{Asner}. The gradual increase along a correlated segment represents a loss of energy to cyclotron radiation. The discrete steps in the frequency come from shallow-angle scatter off the target material. Reprinted figure with permission from D.M. Asner, {\it et al.}, {\it Phys. Rev. Lett.}, {\bf 114}, 162501 (2015). Copyright (2015) by the American Physical Society.}
\label{fig:waterfall}
\end{figure}

Consecutive maxima in the power spectra can be correlated to create a track that represents the energy of the electron. The resulting plot shows both a gradual energy loss to the cyclotron radiation as well as discrete energy changes stemming from shallow-angle scatters with the target atoms.
Fig.~\ref{fig:waterfall} shows an example spectrogram from Asner {\it et al.},~\cite{Asner}.
Reconstruction of the initial frequency of the electron can be transformed into an ejected-electron energy via the equation

\begin{equation}
\label{eq:FreqEnergy}
    f = \frac{eB}{2 \pi \left ( m_e + T/c^2 \right )}
\end{equation}

\noindent
\noindent
where $e$ is the charge on the electron, $B$ is the magnetic field, $T$ is the electron's kinetic energy, $c$ is the speed of light, and $m_e$ is the mass of the electron.

From Eq.~\eqref{eq:FreqEnergy} we see that the resolution of the energy reconstruction is related to the accuracy with which we measure the starting frequency of the track, and there are a number of parameters which contribute to this accuracy.
In the case of a constant radiofrequency (RF) signal, the resolution of the measurement is inversely proportional to the number of samples over which the Fourier transform is performed; therefore, the longer the measurement time, the finer the frequency resolution.  
Since the frequency of the radiating electron is changing, however, there is a limit to this approach, because applying the Fourier transform to a changing signal will result in a broad feature in the resulting power spectrum.
Another effect is the length of the track, where too short a track will preclude a high-quality fit to the starting point.
A third effect is the power of the signal in the first bin, if the power happens to fluctuate below the noise level or the track begins part way through a sampling window, the measured starting frequency will be systematically high.
Further exploration of these effects on energy resolution are discussed in Section~\ref{s:System Performance}.

The maximum cyclotron frequency occurs at the lowest energy, and for an electron in a 1-T field, this ends up being approximately 27.99\,GHz, while for a 30\,keV electron the cycloton frequency is 26.44\,GHz, and at 100\,keV it decreases to 23.4\,GHz.
A typical digitizer operates at 1\,GS/s, requiring the signal to be down-mixed to below the Nyquist frequency of 500\,MHz to achieve a suitable sample rate.
This is why the frequencies plotted in Fig.\,\ref{fig:waterfall} have been reduced by 24\,GHz in hardware. Multiple stages of down-mixing were utilized in Ref.~\cite{Asner}, and we encourage readers to review that work for additional details.

The readout chain has a bandwidth of a few hundred MHz, requiring the experimenter to tune the mix down to capture signals for the desired electron energy.
To achieve a wider effective bandwidth, the signal would need to be amplified and split into multiple mix-down chains, each with the local oscillator tuned to a different frequency.
An instrumented range of 23.4-27.9\,GHz enabled by parallel down-mixers would be able to simultaneously record electrons between 1 and 100\,keV while retaining single-eV energy precision throughout, though that would require a digitizer sampling at a minimum of 9~GS/s.

As a final comment on the apparatus design, we note that, depending on the configuration of the decay chamber, the recorded signal can have systematic effects that can lead to misidentification of the main microwave carrier signal~\cite{Esfahani2}, leading to an error in the energy reconstruction.
These effects must be accounted for in the analysis stage.

%
%

\section{System Performance}
\label{s:System Performance}

In this section we discuss optimization of the experimental setup.
We review choices of target material and density, the effect of electron trap depth, an analysis of observed power, and considerations of event rate and pileup rejection.

A sparse noble gas is an ideal target material for a CRES X-ray detector.
Noble gasses are relatively straightforward to recirculate and purify, allowing for the removal of electronegative molecules which can absorb the trapped electron. 
Also, because noble elements tend not to combine into stable molecules, no molecular shifts in the energy spectrum must be considered when deconvolving the raw spectrum (see Section~\ref{s:ExampleCase}).

%
%

\begin{figure}[t]
\includegraphics[width=0.9\columnwidth]{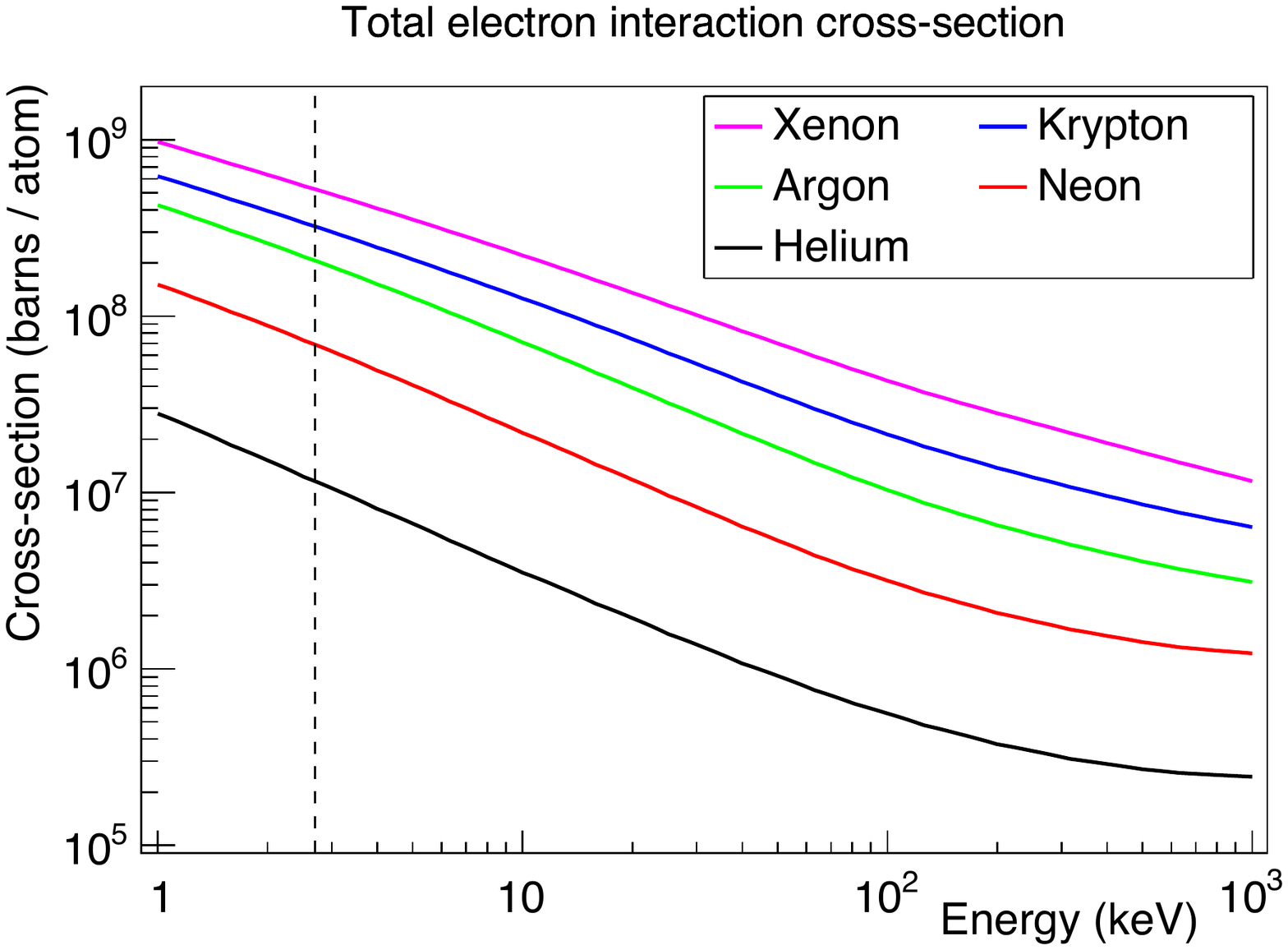}
\caption{Total electron interaction cross-section from the Evaluated Electron Data Library~\cite{Perkins}. The total cross-section was summed over all interaction types available in the database. The dashed vertical line is a visual guide for the 2.7-keV electrons ejected from \iso{55}{Fe} X-rays interacting with an argon atom.}
\label{fig:EEDL}
\end{figure}

The target gas density must be optimized to find a balance between sufficient X-ray interaction targets and the mean free path between electron scatters.
If the electron mean free path is very short, the tracks visible in Fig.\,\ref{fig:waterfall} will be too short for accurate reconstruction of the starting frequency.
Although the tracks visible in that image are on the order of 1\,ms, improved algorithms have been shown to reliably reconstruct tracks down to 200\,$\mu$s~\cite{Esfahani3}.
To obtain a 1/e time between scatters of 200\,$\mu$s, it behooves us to set the target density according to the total electron interaction cross-section for any given target material.
Using the Evaluated Electron Data Library~\cite{Perkins}, we summed the elastic, large-angle elastic, ionization, bremsstrahlung, and excitation cross-sections as a function of electron energy for all the stable noble elements, and used a logarithmic interpolation between available datapoints.
Not all individual interaction cross-sections were tabulated down to 1\,keV, and a linear extrapolation was used to extend all curves to this value, as it proved more stable than a logarithmic extrapolation.
The results are shown in Fig.\,\ref{fig:EEDL}.

The total interaction cross-section can be used to calculate the mean free path, and thereby the mean time between scatters, of the trapped electron as a function of both electron energy and target medium.
If further improvements in event reconstruction can accurately determine starting frequencies with tracks shorter than 200\,$\mu$s, the density of the target can be increased to enhance the data rate.

\begin{figure}[t]
\includegraphics[width=0.9\columnwidth]{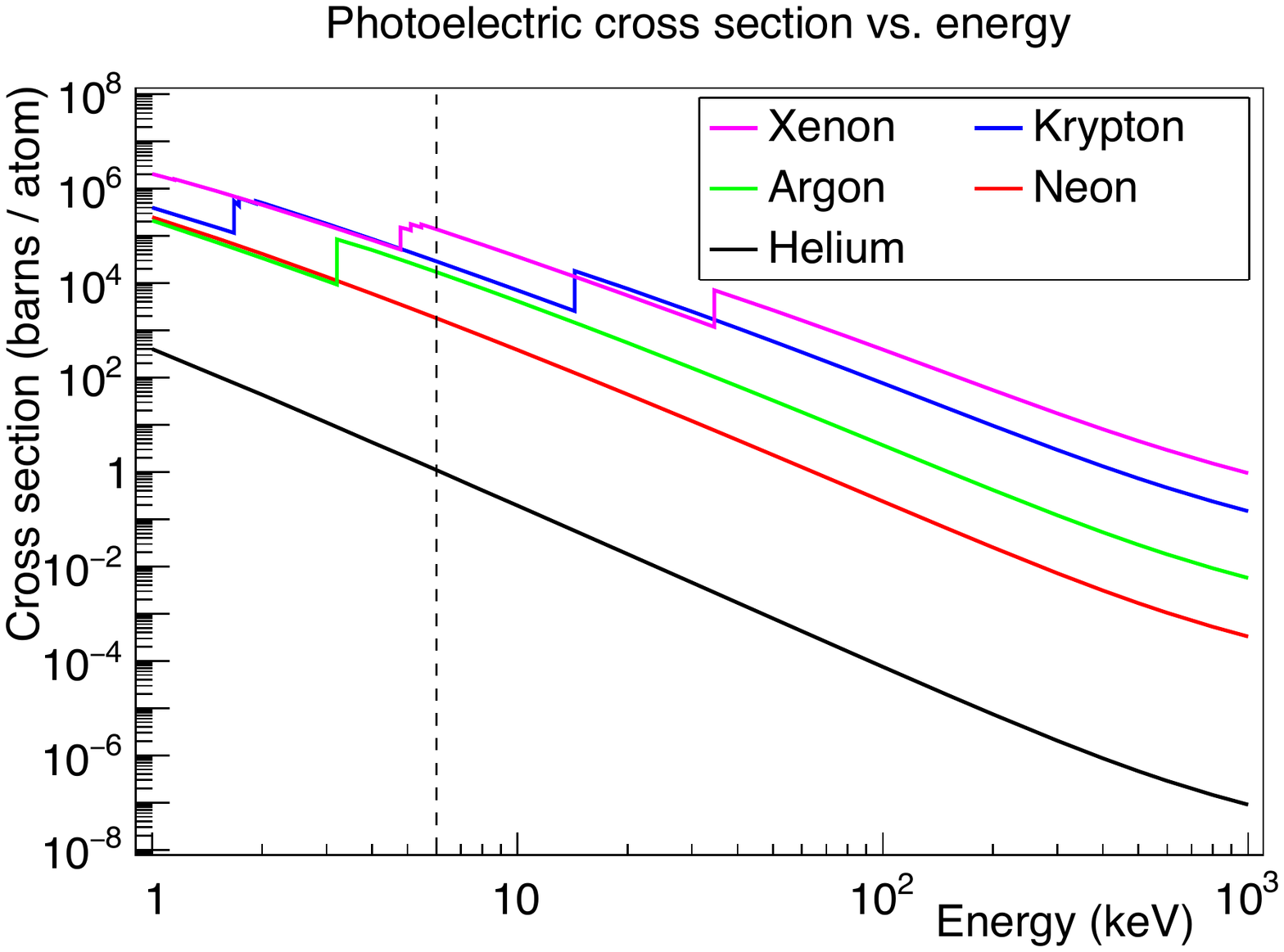}
\caption{Photoelectric cross-section on the stable noble elements. Here, the dashed vertical line is a visual guide for the 6\,keV X-rays from \iso{55}{Fe} decays. Data comes from XCOM~\cite{XCOM}.}
\label{fig:XCOM}
\end{figure}

Higher-Z elements will generally have larger photoelectric interaction cross-sections than lighter elements, but at the cost of greater complexity in the spectral analysis (see Section~\ref{s:ExampleCase}).
The photoelectric cross-section for the stable noble elements is shown in Fig.\,\ref{fig:XCOM}.
At energies near 6\,keV, the photoelectric interaction rate will be greater with a krypton target than an argon target, given an isobaric constraint.
Note, however, that the argon-electron interaction cross-section is smaller than that of krypton-electron interactions, allowing for a greater target density for a given free time between scatters.
The result is that an optimized argon system will have an overall higher event rate than an optimized kyrpton system at this energy.

Next we consider the effect of the depth of the electron trap.
Electrons emitted perpendicular to the main magnetic field ($\theta = 90^\circ$) will spiral in place with no axial motion, while electrons emitted parallel to the magnetic field ($\theta = 0^\circ$) are guaranteed to escape the trap.
For small trapping fields, Esfahani {\it et al.} 2019 give the trapping angle as

\begin{equation}
\label{eq:TrapAngle}
\theta_{\mbox{trap}} = \mbox{sin}^{-1} \left ( \sqrt{1 - \frac{B_{\mbox{trap}}}{B_{\mbox{main}}}} \right )
\end{equation}

The higher the trapping field, the more electrons will be trapped.
There is a point of diminishing returns, however, related to the power emitted by the spiraling electron, given by

\begin{equation}
\label{eq:LarmorFormula}
P = \frac{\pi}{\epsilon_0} \frac{2 e^2}{3 c} f_0^2 \frac{\beta^2 \mbox{sin}^2 \theta}{1-\beta^2}
\end{equation}

\noindent
where $f_0$ is the cyclotron frequency as the electron kinetic energy goes to 0, and $\beta = v/c$. 
The power radiated is maximal when $\theta = 90^\circ$, and zero when the electron is emitted parallel to the main B field.
It is possible that an electron, though trapped within the active volume according to Eq.\,\eqref{eq:TrapAngle}, nonetheless results in no observable signal because the radiated power is too small relative to the system noise to reconstruct the tracks in the power spectrum.
This can also be the case when the design of the RF chain before the first amplification stage has significant losses, reducing the observed signal-to-noise (S/N) ratio. In this situation, the performance of the system would not be improved by increasing the trapping field unless the RF background and electronics noise were reduced to increase the statistical significance of the lower-power signal. The RF background is thermally dominated, and therefore a lower temperature is more desirable, but too low a temperature will greatly reduce the vapor pressure of the gas, and the atoms will simply stick to the wall, eliminating the target atoms from the usable volume

\begin{figure}[t]
\includegraphics[width=0.9\columnwidth]{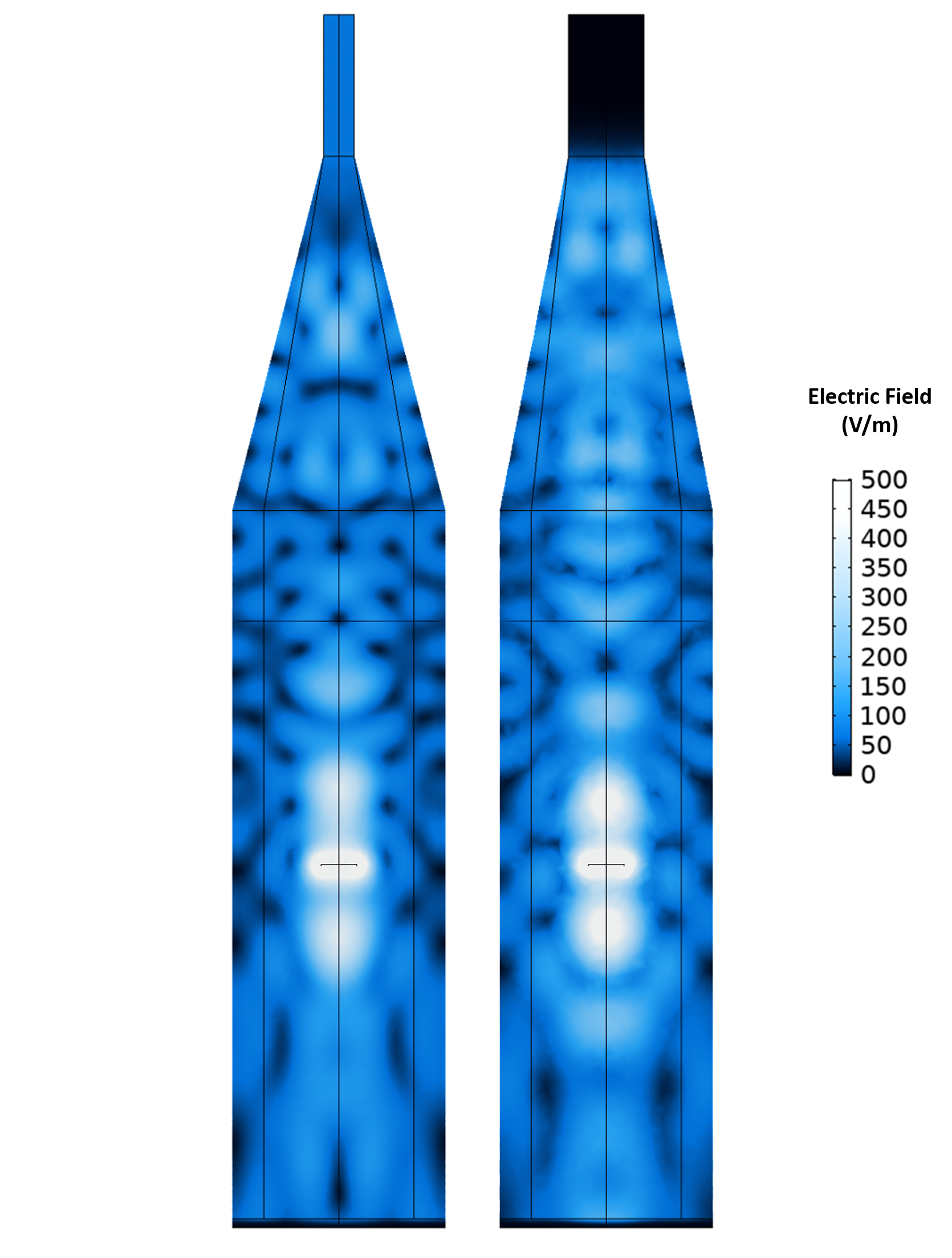}
\caption{Magnitude of the electric field in the target chamber produced by a dipole antenna in the central position with the Z orientation, left, and Y orientation right.}
\label{fig:COMSOL}
\end{figure}

The pre-amplification losses are highly sensitive to the design of the target chamber and the antenna used to capture the RF signal. COMSOL\textsuperscript{\textregistered} simulations were performed to estimate the losses in these components based on the description in Section\,\ref{s:Apparatus}. Two dipole antennas were used to approximate the radiation pattern of the cyclotron radiation with their orientation such that they form a plane perpendicular to the magnetic field. To approximate the RF losses from the X-ray source, the open end of the target chamber was modelled using perfectly matched layers which prevent reflections. A frequency domain analysis at 27\,GHz with the antenna positioned at a range of axial and radial positions acting as port 1, and a WR42 waveguide as port 2 was performed allowing for S-parameters to be extracted. An example solution for the central antenna position is shown in Fig~\ref{fig:COMSOL}. Taking a mean of the positions sampled gives an approximate power collection efficiency of 36\%, though this could be increased through an improved design for the target chamber and antenna horn.

In our COMSOL exploration of the waveguide properties of the  cylinder and antenna, we discovered that there was a notable loss in RF efficiency when the frequency grew too high.
In a 1\,T field, a 2.7\,keV electron would emit cyclotron radiation at 27.8\,GHz instead of 27.0\,GHz.
Our motivation for considering this specific energy is given in Section~\ref{s:ExampleCase}.
At this higher frequency, the RF S/N ratio drops by more than a factor of 2.
In our conceptualization, therefore, we reduced the bulk magnetic field to 0.971~T, putting a 2700\,eV electron at 27.0\,GHz.

\begin{figure*}[t]
\includegraphics[width=\textwidth]{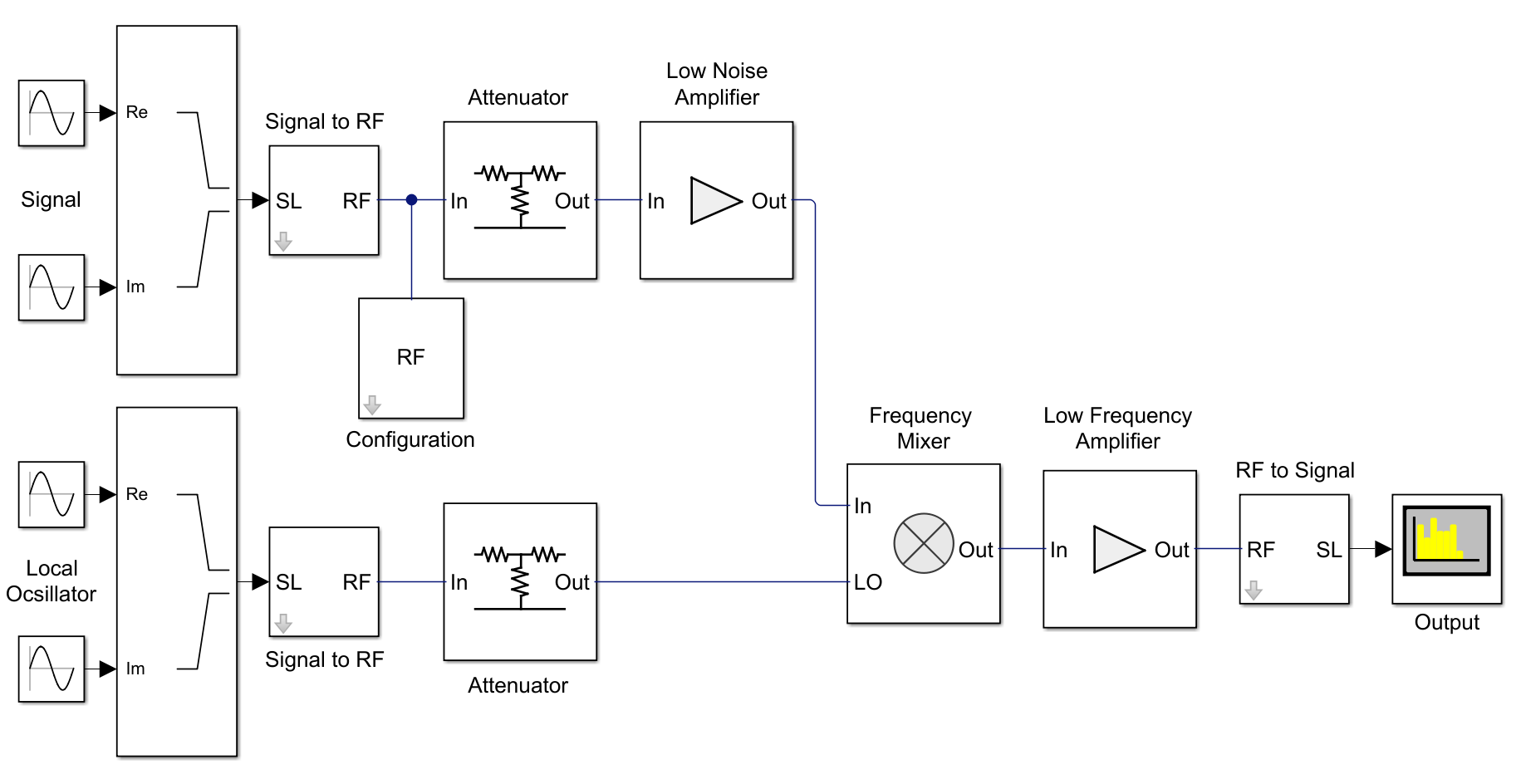}
\caption{Matlab Simulink model of the receiver chain. Complex Simulink signals are defined by the real and imaginary components of the signal coming from the target chamber and the local oscillator for the frequency mix down. These signals are converted for use with the RF Blockset. The performance of elements are taken from appropriate commercial components (multiples of components and filters are not shown in this schematic for simplification).}
\label{fig:MatlabLayout}
\end{figure*}

\begin{figure}[t]
\includegraphics[width=0.9\columnwidth]{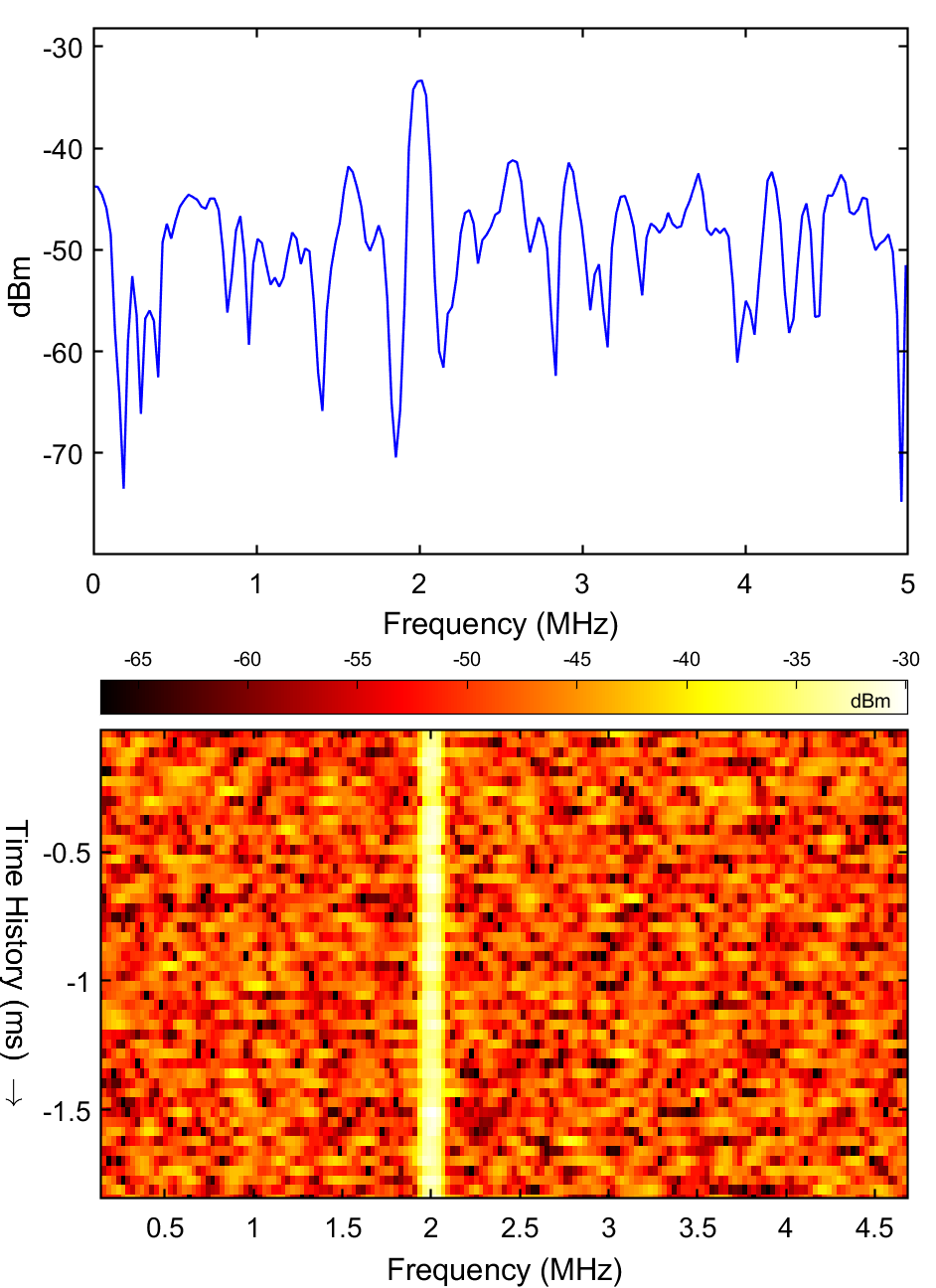}
\caption{Results from the Matlab Simulink receiver chain simulation. \textit{Top:} Sample spectrum with a 100\,kHz resolution of a signal mixed down to 2\,MHz. The signal peak is 7.85\,dB above the peak noise and 16.5\,dB above the mean noise power. \textit{Bottom:} 2ms spectrogram of the sample data with a 100\,kHz frequency resolution and 10\,$\mu s$ time resolution. The 2\,MHz signal is easily visible above the noise. It is worth noting that the expected signal will change frequency which is not shown here. }
\label{fig:SampleSpectrum}
\end{figure}

To estimate the efficiency of track detection the signal power needs to be equal to or greater than the noise power, and preferably significantly above it.
Since the target chamber is well shielded from external RF noise sources, the dominant background will be thermal in nature.
If the target gas is held at 77\,K this would be $10^{-21}$~W/Hz.
Using Eq.~\eqref{eq:LarmorFormula} we calculate that a 2700-eV electron in a 0.971-T field will emit cyclotron radiation at $1.6 \times 10^{-16}$\,W and 27\,GHz.
We used Matlab Simulink to calculate the expected signal on top of the thermal noise, with details of the simulation setup shown in Figure~\ref{fig:MatlabLayout}.
A 200-nV sinusoid was used as the input signal, which was passed through a 6\,dB attenuator to model the power collection efficiency and allow the model to include thermal contributions. The local oscillator was also passed through a 1\,dB attenuator to model the effect of noise in the down-mixer which is used to bring the signal frequency below the Nyquist frequency of the digitizer. The amplifier gain and noise contributions were taken from suitable, commercially available systems.  
By applying a 77-K thermal model, it was shown that S/N is approximately 30 for an unscattered electron (Fig.~\ref{fig:SampleSpectrum}).

The visibility of a peak in the power spectrum, and therefore the ability to reconstruct a track in the spectrogram, depends on the S/N ratio.
The noise power depends in part on the frequency binning, which in turn is partly a function of the time window used for the Fourier transform.
There are two effects with respect to the width of this time window that degrade the S/N ratio.
The first is that the noise is distributed across the frequency spectrum but the signal is restricted to a finite frequency band, and if the frequency bins are significantly wider than the signal band, the noise contribution in that bin is also increased.
The second is the changing frequency of the electron.
If the RF signal were truly static, then a longer time window would be preferable.
Because the frequency is increasing, however, performing a Fourier transform on a very wide time window will lead to a broad plateau in the power spectrum rather than a peak making track construction difficult.
Even beyond these considerations regarding a clear signal above noise is the ability to construct a starting frequency in a track, which requires fitting a line to the signal in the spectrogram.
We note that the wider the time window, the narrower the frequency bins can be, which would lead to an optimal ratio of time bin width and frequency bin width. Though an optimization of the time window is beyond the scope of this work, we calculate the size of the RF signal after processing it through a readout chain designed with a reasonable set of assumptions, and using a time window of 10 $\mu$s, comparable to the 40-$\mu$s time window of Anser {\it et al.}~\cite{Asner}.

\begin{figure}[t]
\includegraphics[width=0.9\columnwidth]{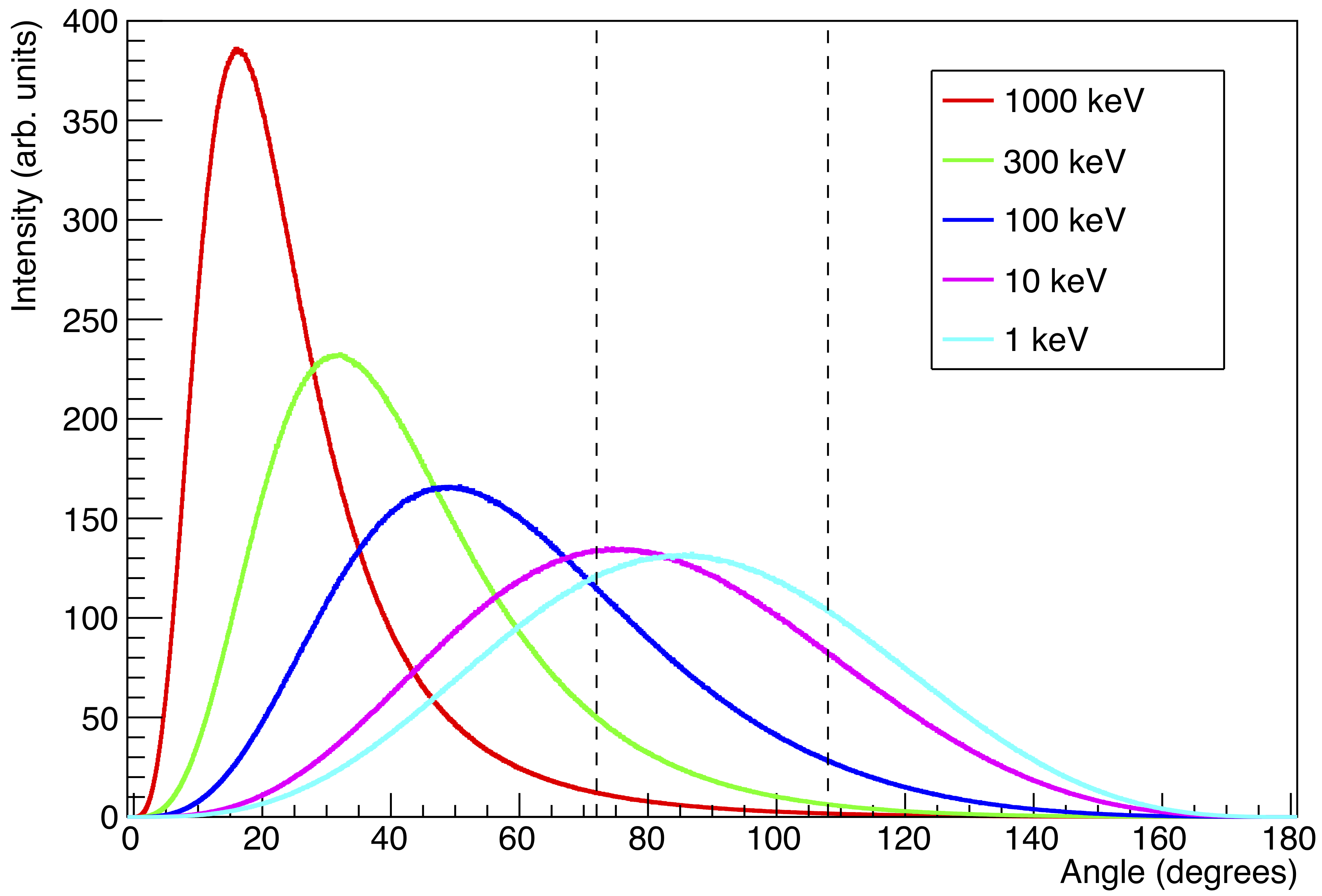}
\caption{Angle of emitted electron from photoelectric interaction with respect to incoming photon. We have integrated over the $\phi$ angle, and thus with respect to the schematic shown in Fig.~\ref{fig:schematic}, this figure represents the intensity of electron emission relative to the main magnetic field. The vertical lines are a visual guide to estimate trapping efficiency between $72^\circ$ and $108^\circ$, where the radiated cyclotron power remains above 90\% of maximum.}
\label{fig:PEAngle}
\end{figure}

Next we consider the bias in electron trajectory following a photoelectric interaction.
Fig.\,\ref{fig:PEAngle} shows the distribution in the emission angle of the electron with respect to the main magnetic field, assuming the incident X-rays are parallel to said field.
A trap depth of 95\,mT on a 0.971-T main field will trap electrons between $72^\circ$ and $108^\circ$.
At the extremes of this range, the power radiated by a trapped electron will be at 90\% of maximum, and therefore still likely to be properly reconstructed, relative to an electron emitted at $90^\circ$.
In this angular acceptance range, an incoming 1-keV photon will have a 42\% probability of ejecting a trapped electron.
At 1\,MeV, the forward momentum of the ejected electron population reduces that probability to about 2\%.
Thus for an experiment optimized to detect photons greater than 100\,keV, placing the source to the side of the active volume, rather than off one end, may increase the overall detection efficiency. 

Related to the electron trajectory is the effect of wall collisions.
If an electron that would otherwise be trapped is emitted within two cyclotron radii of a physical surface, the electron will impact and be absorbed by that surface before a full cyclotron revolution.
Thus the effective radius of the system is reduced by two cyclotron radiii.
For large systems, the remaining volume will be sufficient to perform a measurement but there is a lower bound on the radius of the system of approximately 2.5\,mm, below which the electron is almost guaranteed to be absorbed, precluding observation of any RF signal.

Finally we turn our attention to issues of pileup.
Given the very sparse target density (see Section~\ref{s:ExampleCase}), the rate of simultaneous events is anticipated to be near zero.
If, however, two photons were to interact at exactly the same time, there are several parameters that can be used to separate the events:

\begin{itemize}[topsep=0pt,itemsep=0ex,partopsep=1ex,parsep=1ex]
    \item Energy, where two simultaneous ejected electrons will result in two distinct tracks separated in frequency on the power spectrum---thus two electrons at different energies will be interpreted as such, rather than a single electron at the sum of the energies as is typical in ionization and scintillation detectors
    \item Fiducialization, where an extended array of patch antennas could localize multiple signals within the target volume~\cite{Slocum}
    \item Ejection angle, where the emitted electrons will have different angles with respect to the B field, leading to different values of radiated power and therefore different slopes in the time-evolved power spectrum
    \item Scattering, where the two electrons will interact with target atoms at different times, leading to discrete energy losses at discrete times that are separable
    \item S/N ratio, where two ejected electrons have the same energy, angle with respect to the applied B field, in the same fiducial voxel, will exhibit a signal increase of approvimately 3~dB, clearly differentiating the multiple electrons from a single electron
\end{itemize}

%
%

\section{Example case: measurements of argon target with 6-keV X-rays}
\label{s:ExampleCase}

We now work through an example of an experiment to measure the argon atomic energy levels and photoelectric cross-section from low-energy X-rays from the decay of \iso{55}{Fe}.
This source emits photons at energies shown in Table~\ref{tab:FeSource}, while Table~\ref{tab:ArLevels} shows the binding energies of electron shell levels in the argon atom.
For the last two energies in Table~\ref{tab:FeSource} near 6500\,eV, the value of 20.5 is the combined relative intensity.
For simplicity, and given the ambiguity of this tabulation, in the following analysis we assume the full 20.5 relative intensity is entirely in the 6490.45-eV X-rays, with no X-rays emitted at 6535.2\,eV.

\begin{table}[t!]
\centering
\begin{tabular}{|c|c|}
\hline
Energy (eV)  &   Relative Intensity  \\
\hline
5887.65 &   51  \\
\hline
5898.75 &   100 \\
\hline
6490.45 & \multirow{2}{*}{20.5 (combined)} \\
6535.2  & \\
\hline  
\end{tabular}
\caption{X-ray energies and relative intensities from the decay of \iso{55}{Fe}~\cite{LNHB}.}
\label{tab:FeSource}
\end{table}

\begin{table}[t!]
\centering
\begin{tabular}{|c|c|c|c|c|c|c|c|}
\hline
Shell &  K & L$_1$ & L$_2$ & L$_3$ & M$_1$ & M$_2$ & M$_3$ \\
~ &  1s & 2s & 2p$_{1/2}$ & 2p$_{3/2}$ & 3s & 3p$_{1/2}$ & 3p$_{3/2}$ \\
\hline
Energy (eV) & 3206 & 326 &251 & 248 & 29.3 & 15.9 & 15.7 \\
\hline
\end{tabular}
\caption{Electron binding energies for argon, from Ref.\,\cite{KayeLaby}.}
\label{tab:ArLevels}
\end{table}

\begin{figure}[t]
\includegraphics[width=0.9\columnwidth]{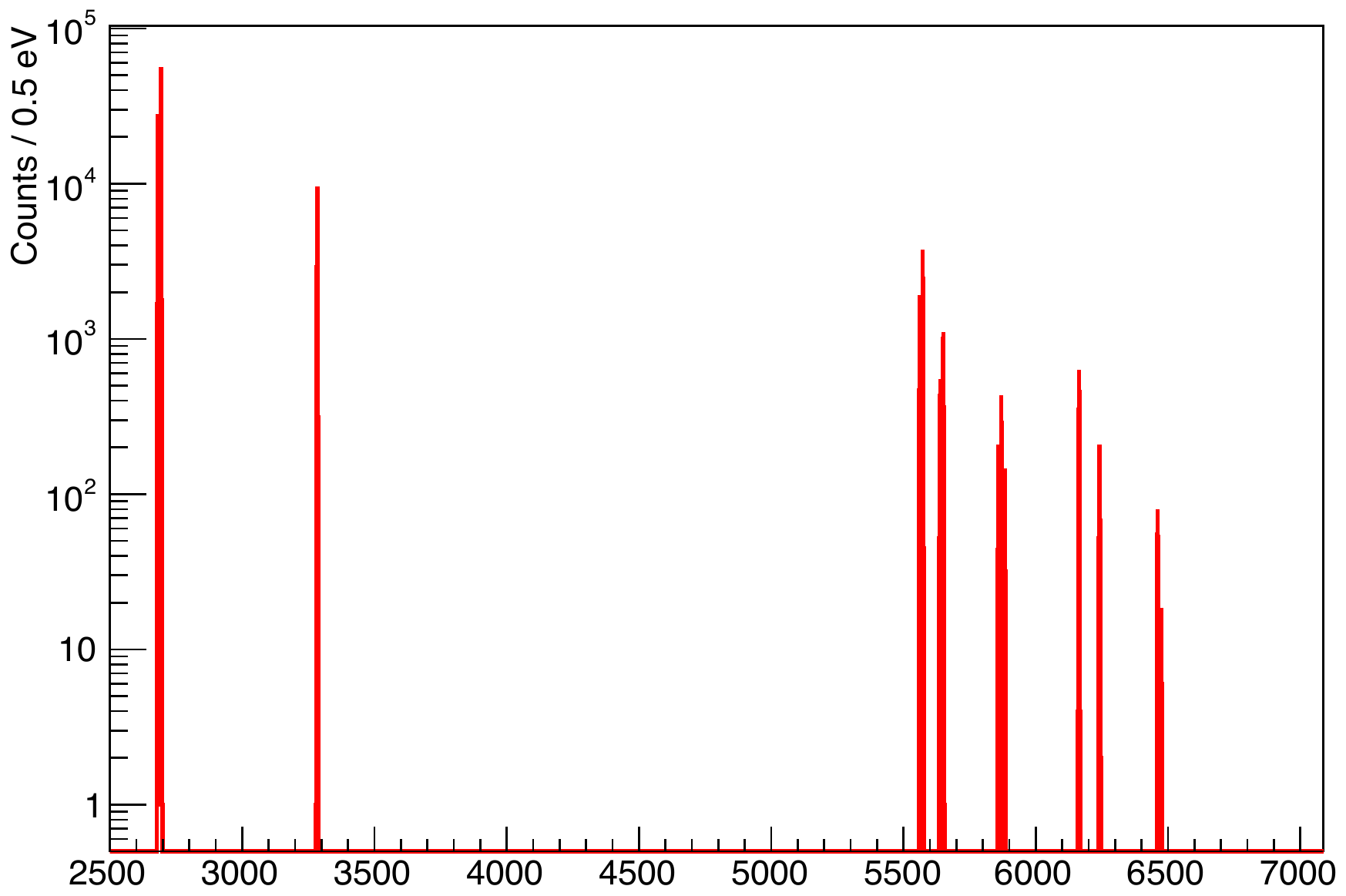}
\caption{Expected observed spectrum from \iso{55}{Fe} decays on an argon target. The assumed resolution is~1\,eV. What appear to be single peaks near 2700, 5570, 5650, and 6470\,eV are doublets separated by 11\,eV, stemming from the X-ray emissions at 5888 and 5899\,eV. Because of the proximity of the energy level differences, there is a triplet of peaks near 5870\,eV.}
\label{fig:FeArSpec}
\end{figure}

Electrons ejected via a photoelectric interaction will have the energy of the incident X-ray minus the binding energy of the shell from which the electron was ejected. In Fig.\,\ref{fig:FeArSpec} we show the expected energy spectrum for \iso{55}{Fe} on an Ar target, taking into account the photoelectric cross-sections given by Scofield~\cite{Scofield}.
For this spectrum, we have assumed a constant energy resolution of 1\,eV across all energies.

The most intense electrons in this scenario are near 2700\,eV, with falling intensities as the energy increases.
This drop in intensity derives primarily from falling photoelectric cross sections on the outer electron shells, although the lower intensity of the 6.5-keV X-rays relative to the 5.9-keV X-rays also contributes to lower electron intensity in the upper energy region.
To optimize the system for the lowest-energy electrons (i.e., those with the greatest interaction cross-section according to Fig.~\ref{fig:EEDL}), we use the total electron interaction cross-section at 2.7\,keV to determine the target density.
The cross-section for argon at this energy is $2.06\times10^{8}$ barns / atom.
A 2.7-keV electron has a velocity of $30.7\times10^6$\,m/s, and if we require a 1/e time between scatters of 200\,$\mu$s, the mean free path must be 6.14~km.
This path length can be achieved by setting the target density to $7.91\times10^9$ atoms / cm$^3$, or $2.23\times10^{-7}$~Torr.
At higher electron energies their cross-section decreases, so the track length of those electrons will be longer and therefore more efficiently reconstructed.



To calculate the event rate, we will assume the active volume is 10\,cm long and 3\,cm in diameter (the same dimensions used in the efficiency calculation shown in Fig.~\ref{fig:COMSOL}).
We used a Monte Carlo approach to calculate the distribution of path lengths in the active volume, where the source was modeled as an extended planar disk 1\,cm in diameter, coaxial with the active volume, 1\,cm away from the trapping coil, and with isotropic emission.
Averaging over all trajectories, including those that did not intersect the active volume, resulted in an average path length of 0.39\,cm.

A more full-featured Monte Carlo calculation would be necessary to incorporate additional physical effects of a real-world design, such as attenuation through a beryllium window for an external source (see Section~\ref{s:Discussion}).
A second effect to consider is the loss of active volume from the finite cyclotron radius, as discussed in Section~\ref{s:System Performance}.

\begin{table}[b!]
\centering
\begin{tabular}{|l|l|l|}
\hline
Variable            &   Description                         &   Value \\
\hline
$A$                 &   Source activity                     &   1.00\,Ci \\
$\left < L \right >$                 &   Average path length                 &   0.390\,cm \\
$\sigma$            &   PE x-sec on K$_{\mbox{1s}}$ shell   &   1.55e4\,b/at \\
$\rho$              &   Target density                      &   7.91e9\,at/cm$^3$ \\
$\epsilon_{emi}$    &   Emission angle efficiency           &   0.421 \\
$\epsilon_{tr}$     &   Track length cutoff efficiency      &   0.368 \\
$\epsilon_{pwr}$    &   Track power efficiency              &   1.00 \\
\hline
\end{tabular}
\caption{Variables used in calculating the event rate on an argon target with an \iso{55}{Fe} source. $L$ comes from a Monte Carlo calculation of the unattenuated path length through the active volume, $\sigma$ comes from Scofield~\cite{Scofield}, $\epsilon_{emi}$ comes from Section~\ref{s:System Performance}, and $\epsilon_{tr}$ is set by establishing the 1/e track length attenuation constant to 200\,$\mu$s. The efficiency for detecting an electron up to $\pm18^\circ$ away from perpendicular, $\epsilon_{pwr}$ is assumed to be close to 100\%, given the S/N ratio of 30.}
\label{tab:OptimizedVariables}
\end{table}

The final rate of events into the lowest-energy peak is given by:

\begin{equation}
\label{eq:EventRate}
\mbox{Rate} = A \left ( 1 - e^{-\left < L \right >\sigma\rho} \right ) ~\epsilon_{emi}~\epsilon_{tr}~\epsilon_{pwr}
\end{equation}

\noindent
where the variable definitions and their values are shown in Table~\ref{tab:OptimizedVariables}.
We use the average path length in the active volume in Eq.~\eqref{eq:EventRate} as a simplification, which is valid given the very long ($\sim$82,000\,km) attenuation length of the X-rays in the target medium.
We have assumed that the trap depth is shallow enough that any trapped electrons will radiate sufficient power to be observed above noise levels at near 100\% efficiency. (see Section~\ref{s:System Performance}).

The final rate of detected K-shell electrons ejected by 5899\,eV X-rays, i.e. counts in the peak at 2693\,eV, is 0.275\,Hz.
We would collect approximately 166,000 events per week into this peak, with approximately half that into the peak at 2682\,eV.
The L-shell peaks and the K-shell ejections from the 6.5\,keV X-rays would have 9,500 to 19,000 counts in this same time frame.
The lowest-intensity peaks would have approximately 120 counts.
With a bin width of 0.5\,eV and a total systematic resolution of 1\,eV, this lowest-intensity peak would have a centroid uncertainty of approximately 0.11\,eV.
This accuracy would allow measurement of the relative energy levels to better than an eV, while the number of counts in each peak would provide a relative measurement of the photoelectric cross-sections, with accuracy ranging from 0.4\% to 15\%.

The spectrum shown in Fig.~\ref{fig:FeArSpec} assumes a constant $\epsilon_{tr}$ across all electron energies.
In reality, because the argon-electron cross-section drops at higher energies, $\epsilon_{tr}$ would be higher, and the number of counts in these peaks correspondingly higher.
When performing a full spectral analysis not only of the peak energies, but intensities, this effect must be incorporated.
Otherwise, the greater number of counts in these higher-energy peaks would be incorrectly interpreted as a higher photoelectric cross-section.


The cyclotron frequency of electrons between 2700 and 6500\,eV is 26.8 to 27.0\,GHz for a 0.971-T field.
Thus if 26.7\,GHz is subtracted from the signal using a down-mixer, and the system's bandwidth was at least 200\,MHz, the full spectrum could be recorded with frequencies between 100 and 300\,MHz.

Once a spectrum is obtained, it must be deconvolved to reconstruct the original X-ray emissions.
The lowest-energy peak in the spectrum needs to be associated with higher-energy peaks from any L- and M-shell ejections resulting from the same energy of incoming photon. Establishing this correlation will increase the significance of the primary peak.
The lowest-energy peak should then be augmented by the binding energy of the K-shell electron, and normalized by the photoelectric cross-section to recover the original X-ray energy and intensity. The next-lowest unassociated peak will then be identified, with its own associated L- and M-shell correlations, and augmented in the same manner as the first peak.
Each peak will therefore be identified in turn.
Thus the full spectrum of the original X-ray energies and intensities is reconstructed.
The heavier target atoms have more shells from which electrons can be ejected, leading to a more complicated spectrum with a greater number of associated peaks, although the analysis approach presented here is still applicable.

We end this section with a concept for a systematic test, where the \iso{55}{Fe} source is replaced by a \iso{49}{V} source.
This source has a comparable half-life, and emitted X-ray spectrum to the \iso{55}{Fe}, and provided the argon target properties are consistent, differences in the resulting spectra can be attributed directly to the use of a different source.
A custom radioactive source might incorporate both \iso{55}{Fe} and \iso{49}{V}, removing many systematic differences between data acquisition runs.

%
%

\section{Discussion}
\label{s:Discussion}

The energy levels of all the noble elements can be measured using the approach outlined in this work.
The accessibility of those levels is directly related to the incoming photon energies, i.e., a K-shell electron from a Xe atom can be ejected only if the incoming photon is in excess of 35\,keV.
The accuracy of the measurement comes from both the instrumental resolution and the precision of the incoming photons.
Given the energy and precision constraints, a good choice for irradiation usable for all the noble targets might be the 60\,keV gammas from the decay of \iso{241}{Am}.

We now consider the event rate between the lightest and heaviest stable noble elements to delineate the extremes of system performance.
The photoelectric cross-section at 60\,keV is seven orders of magnitude greater for xenon than helium.
The electron interaction cross-section, though, is two orders of magnitude greater.
This simplistic consideration would give an event rate in an optimized xenon system $10^5$ times that of the event rate in the helium, but with a big caveat: the population of the M- and N-shell peaks from the xenon are also attenuated by several orders of magnitude, in a manner similar to what is shown for the L and M shells for argon in Fig.~\ref{fig:FeArSpec}, while the helium target only has a single shell to characterize.
A full design consideration of the two targets and an \iso{241}{Am} source might show a surprising parity between the required run times.

As previously mentioned, the system can be optimized for any particular X-ray energy and target medium.
The optimization process starts with determining the shortest acceptable path length of the trapped electron, based on the total interaction cross-sections in Fig.~\ref{fig:EEDL}.
This path length then determines the target density.
Given the very long attenuation length of the sparse targets, the event rate will scale directly with the volume, although the observation of the CRES signal in a large volume will require alternative readout hardware, such as that described for the Free-Space CRES Demonstrator for Phase III of Project 8~\cite{Oblath}.
In addition to optimizing the target, the RF properties can also be optimized for any given electronics chain or digitizer speed via appropriate selection of the applied magnetic field as well as the frequency of the down-mixing.

As a final concept, this approach to high-precision X-ray spectroscopy can be expanded to characterization of external materials.
If the source in Fig.~\ref{fig:schematic} is replaced with a beryllium window, a target outside that window can be induced to emit X-rays via irradiation from various sources, e.g., from \iso{241}{Am} similar to measurements performed by Turhan {\it et al.}~\cite{Turhan}), X-rays from a broad-spectrum beam as described by Higley {\it et al.}~\cite{Higley}, or electrons from a sufficiently intense electron gun.
Once the energy levels of the gas in the active volume are determined to high precision, the energy levels of external targets can also be measured and potentially used for identification of unknown samples.
This interrogation of unknown samples may prove particularly effective in observing the K-shell interactions of low-Z trace elements in a background of L-shell interactions of medium-Z bulk elements~\cite{Wachulak}.

%
%

\section{Summary}
\label{s:Summary}

In this work we have presented a concept for an X-ray detector based on cyclotron radiation emission spectroscopy, giving heightened priority to resolution over other detector metrics such as intrinsic efficiency.
We have explored the detector choices and setpoints within the context of measuring the energy levels of the argon atom, although the concepts presented here may be incorporated into other experimental needs.
As far as the argon conceptualization goes, we explore the issues surrounding measuring the energy levels and photoelectric cross-sections for the shells of not just argon, but all the stable noble elements.
The apparatus is anticipated to have eV-scale resolution, thereby testing the theoretical calculations underpinning many physics experiments to unprecedented accuracy.
We have described experimental considerations to optimize a target, with dependencies on photon energy, target density, photoelectric cross-section, electron interaction cross-section, electron emission angle distribution, active volume and source configuration, electron trapping depth, and signal in the power spectrum above the noise floor.
With high-resolution measurements of the noble element energy levels this work can be extended to characterize external materials if the internal source were replaced with a low-Z window and the external material is illuminated by photons or electrons of a compatible energy.

%
%

\section{Acknowledgements}
\label{s:Acknowledgements}

We would like to acknowledge the Project 8 collaboration, who have collectively pioneered the CRES concept. We would also like to thank Neal Carron, who took great pains to update the EEDL documentation, even though he was not on the author list. Lawrence Livermore National Laboratory is operated by Lawrence Livermore National Security, LLC, for the U.S. Department of Energy, National Nuclear Security Administration under Contract DE-AC52-07NA27344. Work supported by LLNL LDRD, tracking \#20-LW-056. LLNL-JRNL-790159.


\begin{thebibliography}{99}
\bibitem{KayeLaby}Kaye and Laby Tables of Physical and Chemical Constants, National Physics Laboratory, United Kingdom. This database is no longer actively maintained, but is still available through web archive sites such as web.archive.org. 
\end{thebibliography}
\end{document}